# Drug repurposing prediction for COVID-19 using probabilistic networks and crowdsourced curation

David J. Skelton[1], Aoesha Alsobhe[1], Elisa Anastasi[1], Christian Atallah[1], Jasmine E. Bird[2], Bradley Brown[1], Dwayne Didon[1], Phoenix Gater[1], Katherine James[3], David D. Lennon Jr[1], James McLaughlin[1], Pollyanna E. J. Moreland[1], Matthew Pocock[1,4], Caroline J. Whitaker[4], Anil Wipat[1,*]

[1]School of Computing, University of Newcastle upon Tyne, Newcastle upon Tyne, United Kingdom

[2]School of Natural and Environmental Sciences, University of Newcastle upon Tyne, Newcastle upon Tyne, United Kingdom

[3]Department of Applied Sciences, Northumbria University, Newcastle upon Tyne, United Kingdom

[4]Turing Ate My Hamster Ltd, Newcastle upon Tyne, United Kingdom

*Corresponding author

E-mail: anil.wipat@newcastle.ac.uk (AW)

# Abstract

Severe acute respiratory syndrome coronavirus two (SARS-CoV-2), the virus responsible for the coronavirus disease 2019 (COVID-19) pandemic, represents an unprecedented global health challenge. Consequently, a large amount of research into the disease pathogenesis and potential treatments has been carried out in a short time frame. However, developing novel drugs is a costly and lengthy process, and is unlikely to deliver a timely treatment for the pandemic. Drug repurposing, by contrast, provides an attractive alternative, as existing drugs have already undergone many of the regulatory requirements. In this work we demonstrate an approach to drug repurposing using a combination of probabilistic and semantically rich networks. This combination, together with clustering algorithms, facilitates both the use of network algorithms and distributed human curation, to search integrated knowledge graphs, identifying drug repurposing opportunities for COVID-19. We demonstrate the value of this approach, reporting on eight potential repurposing opportunities identified, and discuss how this approach could be incorporated into future studies.

# Introduction

Severe acute respiratory syndrome coronavirus two (SARS-CoV-2), the virus responsible for the coronavirus disease 2019 (COVID-19) pandemic, was discovered in Wuhan, China, in December 2019 [1]. SARS-CoV-2 is a member of the betacoronaviruses, a genus of enveloped positive-sense single-stranded RNA viruses. The symptom profile of COVID-19 infections vary, with a dry cough and fever most often reported [2] and resulting in a mild illness in most cases. However, severe illness has been reported as occurring in as many as 20% of laboratory-confirmed infections [3], depending on population demographics and country-specific testing protocols, with symptoms including acute myocardial injury [4] and acute kidney injury [5]. Two other highly pathogenic coronaviruses are known [6]: SARS-CoV-1, responsible for the SARS outbreak in 2002-2004, and the Middle East respiratory syndrome (MERS) coronavirus, responsible for the MERS outbreak in 2012. Unlike SARS-CoV-1 and MERS-CoV, however, SARS-CoV-2 is widely reported to cause asymptomatic carriage [7] in some individuals, which has contributed to the rapid, global spread [8], and subsequent radical, world-wide restrictions on daily life to control transmission. The disease now represents an unprecedented global health challenge.

## Mechanism of infection

Understanding of the mechanism of pathogenesis is a piecemeal process. One process that is particularly well understood is the entry of SARS-CoV-2 into human cells, which occurs via two major methods: TMPRSS2-dependent entry and TMPRSS2-independent entry [9]. The first method relies on two human transmembrane proteins, ACE2 and TMPRSS2. The spike proteins of SARS-CoV-2, which mediate coronavirus entry to human cells, have been shown to have strong affinity for the human protein ACE2, which is highly expressed on the surface of different cell types, including lung epithelial cells and type II alveolar cells [10]. Once the viral spike proteins have bound to ACE2, localised TMPRSS2 can activate the viral spike proteins and cleave the cytoplasm-facing domain of ACE2. This mechanism causes the virus' membrane to fuse with the human cell's membrane, allowing the viral genomic RNA to enter the cell, where it can replicate and facilitate the production of new virions.

In the absence of localised TMPRSS2, SARS-CoV-2 can enter human cells expressing ACE2 via another mechanism. In this process, the viral spike proteins bind to ACE2 and the ACE2 receptor protein recruits clathrin, which initiates endocytosis of the virus. Usually, following endocytosis, proton pumps transport $H^+$ ions into endosomes, lowering the internal pH, allowing the endosome to begin the process of becoming an endolysosome and breaking down anything within. However, the pH-dependent enzyme cathepsin L present in endosomes are used by SARS-CoV-2 to activate the viral spike proteins, which allow the virus to fuse with the endosomal membrane and release its genomic RNA into the host cell in preparation for replication and production of more virions.

## Repurposing drugs for the treatment of COVID-19

Due to the rapid increase in COVID-19 cases, a large amount of research into the disease pathogenesis and potential treatment has been carried out in a short space of time. Given the rapid onset of the pandemic, there is an urgent need to identify new drugs that are able to treat the disease and improve patient outcomes [11,12]. Developing novel drugs and taking them to market typically takes many years, so the traditional pathways are unlikely to provide a timely treatment for the current pandemic. The time pressures for new treatments make drug repurposing particularly attractive. Drugs approved for one indication (disease) may also happen to be efficacious for another. If likely candidates for repurposing can be identified, the time to approval for a drug can be reduced to just a few months by drawing from the pool of the many thousands of drugs that have already been through the regulatory approval process, and are therefore well-characterised. Identifying high quality drug repurposing candidates is challenging. Many discoveries are serendipitous, for example from observations in clinics and trials [13]. However, with the advent of network approaches to studying biology at a systems level, new computational approaches have emerged that allow a more systematic approach to drug repurposing [14–16]. Networks are a convenient method to represent drugs, genes, proteins, diseases, etc., and the relationships between them. They support human investigation through visual exploration, while at the same time being amenable to computational analysis, enabling automated and systematic analysis.

Drug repurposing algorithms derive new relationships between drugs, targets, diseases, genes etc., based on existing knowledge of these relationships [15], These new relationships can lead to hypotheses about novel repurposing opportunities and the evidence for them can be reviewed and then hypotheses tested in the laboratory and clinic. Computational drug repurposing prediction and human curation both require the availability of a variety of data sources that describe the molecular processes in a cell and their interaction with drugs. These data are typically distributed in a range of different data sources and databases and may be integrated after they have been represented as networks to form an integrated network or knowledge graph [17–19]. Repurposing algorithms that operate over the data can vary from simple pattern matching or network module analysis, through to the application of artificial intelligence [20–24]. The output of these algorithms are systematic computational predictions and therefore need human curation, with a range of expertise, to assess their value. Therefore, the computational prediction of repurposable drugs is usually applied in an iterative fashion, coupled with the input of a human expert who is able to verify and generate hypotheses about new repurposing opportunities.

The need for new drugs to treat COVID-19 has resulted in a large number of recent studies [25–41]. Antiviral drugs have been the most obvious target since some of these have been shown to be active against SARS. For example, drugs such as Elbasvir are under scrutiny to this effect [42]. Other antiviral drugs, such as Remdesivir, originally employed for the treatment of Ebola, have recently been approved for the treatment of COVID-19 [43]. However, there is still a pressing need for more candidate drugs, especially those addressing the abnormal immune response seen in some patients [43,44]. The wider search for repurposable drugs remains a field under active development. Most of these studies are computational in nature and have resulted in many potential new drug candidates. These studies have employed a variety of strategies including network based analyses. Some of these candidates are now undergoing clinical trials to test their effectiveness [34].

In this work we used a combination of network algorithms and human curation to search integrated knowledge graphs for drug repurposing opportunities for COVID-19. We aimed to highlight potential repurposing opportunities that are possibly novel in comparison to those published so far. We also introduce some newer features to our analytical process in an effort to improve accuracy by incorporating measures, which take data quality into account, and to speed up data verification by facilitating multiple curation.

Many studies, for example that of Kumar 2020 [36], have employed protein interaction networks to study the relationships between COVID-19 proteins, the proteins they interact with, and the molecular interaction networks in the cell. However, data quality is an important factor in the assignment of protein interactions, and has received little attention in studies to-date. Instead of using a standard protein interaction graph, we developed a probabilistic functional interaction network for human proteins that uses highly curated interactions from BioGrid as a gold standard to produce a weighted protein interaction network, where the weight indicates the confidence of a functional interaction. We then mapped onto this human probabilistic function integrated network (PFIN) the interactions with COVID-19 proteins, discovered experimentally in Gordon *et al.*, 2020 [27].

The process of curating computational predictions, or manually analysing networks for new opportunities is a large-scale, time consuming task. The division of labour through network partitioning and crowdsourcing are promising approaches to more rapid verification of results [45,46]. However, one of the challenges facing this approach is the assignment of appropriate sections of a complex network to each curator. Since we were working with PFINs that included weighted edges, we were able to employ a clustering algorithm to partition the probabilistic network into functionally related subgraphs. Those subgraphs containing COVID-19 protein interactors were then enriched with information from previously integrated networks to add relationships to drugs, diseases, genes, pathways and so on. Each subgraph then represents a unit of the network that was systematically assigned to a curator for analysis. In this fashion, a large scale task is broken down and becomes faster with more curators participating in the exercise. Here we present details of our approach to the development and partitioning of the knowledge network and the resulting predictions for potentially repurposable drugs.

## Methods

This work extends approaches developed by Mullen *et al*. [47]. In brief, a semantic knowledge graph was constructed in Neo4j, comprising data about existing, approved drugs. The data sources contributing to the knowledge graph are shown in Table 1.

| Data source | Date obtained | Nodes contributed | Edges contributed |
|---|---|---|---|
| UniProt [48] | 2020-02-11 | Proteins | Gene -[encodes]- Protein |
| DrugBank [49] | 2020-02-11 | Drugs | Drug -[has target]- Protein |
| Monarch Disease Ontology (MONDO) [50] | 2020-02-11 | Disorders | Disorder -[is subset of]- Disorder |
| Drug Central [51] | 2018-08-26 | | Drug -[has target]- Protein  Drug -[has indication]- Disorder |
| OMIM [52] | 2020-02-07 | | Gene -[associated with]- |

|  |  |  |  Disorder |
| --- | --- | --- | --- |
| DisGeNET [53] | 2019-12-02 |  | Gene -[associated with]- Disorder |
| NCBI Gene Info [54] | 2020-02-11 | Genes |  |

Table 1: Databases integrated as part of this study.

Using the Neo4j knowledge graph, two approaches were taken to identifying drug repurposing candidates. In the first approach, a search of literature on COVID-19 and SARS-CoV-2 was carried out to identify current knowledge on mechanisms of disorder pathogenesis, including genes and proteins in *Homo sapiens* and SARS-CoV-2 that are involved in this process. Based on the results of the literature search, semantic queries were devised to explore the local neighbourhood around identified concepts of interest in the knowledge graph.

In the second approach, a PFIN was created from BioGrid [55] v181. A PFIN is a network in which the nodes (in this case, genes) are linked via weighted edges, with the weight representing the confidence that the two genes functionally interact. The PFIN was constructed using a method devised by Lee *et al.* [56]. In brief, BioGrid v181 was partitioned into high-throughput (HTP) datasets and low-throughput (LTP) datasets, with low-throughput datasets being defined as datasets containing fewer than 100 interactions. By considering the LTP datasets as a gold-standard, log likelihood scores (LLS) for the high-throughput datasets were calculated using an approach as described by Lee *et al.* [56],

$$LLS = \ln\left(\frac{P(L|E)\,/\,{\sim}P(L|E)}{P(L)\,/\,{\sim}P(L)}\right)$$

where, P(L|E) and ~P(L|E) represent the frequencies of linkages (L) observed in dataset (E) between genes in the gold-standard, and absent from the gold-standard, respectively, and, P(L) and ~P(L) represent the prior expectation of linkages between genes in the gold-standard, and absent from the gold-standard, respectively.

Datasets with a LLS of 0 or lower were discarded, and the remaining datasets were sorted in order of LLS, and integrated using the following formula:

$$WS = \sum_{i=1}^{n} \frac{L_i}{1.1^{i-1}}$$

where, $L_1$ is the highest LLS score and $L_n$ that with the lowest of a set of *n* datasets. The denominator selected to give more weight to datasets with higher confidence was 1.1, as used in previous work by James *et al.* *[57]*. For visualisation purposes, functional interaction relationships were omitted if the confidence score was lower than 4.336.

The PFIN was clustered using a weighted MCL algorithm with the default parameters and viewed in Cytoscape [58,59]. Next, for the set of *Homo sapiens* proteins identified as interacting with SARS-CoV-2 proteins by Gordon *et al.* [60], clusters from the PFIN containing genes that encode at least one of these proteins were extracted, decorated with edges and nodes from the knowledge base (Fig 1), and returned.

**Fig 1. Cypher query used to decorate genes in PFIN clusters that have at least one gene encoding a protein that interacts with a SARS-CoV-2 protein.** Information obtained from the knowledge base includes (1) proteins encoded by the

genes, (2) drugs that target any proteins in (1), (3) drugs with similarity drugs in (2), (4) disorders associated with the genes in the PFIN cluster.

Due to limitations of the integrated database we constructed, such as a lack of side-effect data and missing annotations for drug actions on their targets, clusters required manual curation to identify promising targets. Thus, this analysis was crowdsourced amongst various contributors. A Trello system was used to coordinate the distribution of clusters, collate information, provide background material and to gather the results of the analysis by curators.

# Results

Results from the cluster analysis from multiple curators were assimilated and integrated and are presented on a per-drug (or per-drug class, where appropriate) basis. Each section below describes (1) the drug / class, (2) the method(s) used to identify the drug / class as a candidate, and (3) an explanation of any relevant literature to support or refute the drug candidacy.

## Theophylline

Theophylline (IUPAC name: 1,3-dimethyl-2,3,6,7-tetrahydro-1H-purine-2,6-dione) is a methylxanthine drug with activities including smooth muscle relaxation and bronchial dilation. Indicated for conditions such as chronic obstructive pulmonary disorder (COPD) and asthma, theophylline exerts its activity through competitive inhibition of type III and IV phosphodiesterase [61]. Clustering of the PFIN revealed a cluster of genes, one of which, *DNAJC11*, encodes the DNA J homolog subfamily C member 11 protein (Q9NVH1), which Gordon *et al.* [27] reported as interacting with the SARS-CoV-2 E protein. SARS-Cov2 *E* encodes the envelope protein, E, which plays a role in the structure and maturation of the virus [62]. DNAJC11 is a mitochondrial protein which is required for mitochondrial inner membrane organization which associates with the mitochondrial contact site and cristae organizing system (MICOS) complex [62,63]. The PFIN suggests that *DNAJC11* has a high likelihood of being functionally related to *RIC3*, which encodes the protein RIC3 (Q7Z5B4) -- a protein target of theophylline (Fig 2) [64]. RIC3 promotes functional expression of homomeric alpha-7 and alpha-8 nicotinic acetylcholine receptors at the cell surface and is found in endoplasmic reticulum and the golgi apparatus [65]. The relevance of the functional relationship between *DNAJC11* and *RIC3* is unclear, however the interaction was identified by Affinity Capture-Mass Spectroscopy, which implies a physical binding. It is possible, therefore, that RIC3 complexes with DNAJC11 (although current evidence has them expressed in different organelles) and theophylline could block the binding of DNAJC11 by SARS-CoV-2 E. Interestingly, DNAJC11 has well-conserved orthologues in other species, such as mouse, rat, and dog [66].

**Fig 2: Network showing a connected component containing a gene encoding a protein targeted by theophylline**. This is part of a larger PFIN cluster, but shown as disconnected due to the edge thresholding for visualisation. The red node is Q9NVH1, the pink nodes are genes, and the blue node is theophylline. Q9NVH1 is a high-confidence interactor of the SARS-CoV-2 nsp4 protein.

## Calmodulin inhibitors

Calmodulin is a ubiquitous calcium binding protein which binds to transmembrane proteins such as ACE [67]. As described in the introduction, ACE transmembrane proteins have a distinct role in Covid-19 infection [9]. Calmodulin plays a crucial role in regulating ACE2 presence on the cell surface by binding to the cytoplasmic tail of ACE2 [68]. A potential target for decreasing infection and viral loading could, therefore, be the use of calmodulin inhibitors, acting to decrease the association of the two proteins. Without the binding of calmodulin to ACE2, the cells start to shed the ectodomain of ACE2, leading to decreased expression and catalytic activity [67]. Querying our knowledge graph for approved drugs that inhibit CALM1 (P0DP23), we identified 10 existing, approved drugs that are labelled as inhibitors of calmodulin, indicated for a range of conditions including hypertension, seasonal allergies, and antidepressants (full list in Table S1).

## Angiotensin II

Angiotensinogen is a small peptide hormone which, as part of the renin-angiotensin-aldosterone system (RAAS), is involved in the regulation of blood pressure [69]. Via the action of renin, angiotensinogen is converted into angiotensin I. In turn, angiotensin I, via the action of angiotensin converting enzyme (ACE), is converted into angiotensin II. Critically, ACE2 converts angiotensin II into angiotensin (1-7), an effector of nitric oxide-dependent vasodilation which has been implicated in attenuating acute lung injury [70]. The role of drugs which inhibit the RAAS, such as commonly prescribed antihypertensive agents (e.g., ACE inhibitors and angiotensin receptor blockers) is controversial, with arguments proposed for both harmful and protective effects [71–73]. Given that angiotensin II is, itself, an approved drug, it may have a therapeutic use in COVID-19 via the actions of angiotensin (1-7). Angiotensin II has already been suggested as a possible therapeutic agent, and is currently being investigated (NCT04332666).

## Cathepsin L and TMPRSS2 synergistic targets

Whilst inhibition of the ACE2 receptor may seem an ideal target for treating COVID-19, as it is required for both major mechanisms of viral entry into host cells, it has been shown that ACE2 is essential for repairing damage to lung tissues. Instead, it may be possible to inhibit TMPRSS2, which could lead to most virus particles entering the host cell via endosomes. If cathepsin L is also inhibited, the endosome-enclosed viruses will be unable to escape, allowing the endosomes to progress fully to endolysosomes and destroy the virus particles. In this way, it may be possible to not only prevent entry into the host cell and subsequent replication, but also reduce overall viral load by enabling the body to destroy SARS-CoV-2.

**Table 2. Cathepsin L inhibitors identified in our database.**

| Drug name | Description | Drug groups (according to DrugBank) |
|---|---|---|
| Telaprevir | Antiviral medication used as part of combination therapy in hepatitis C infection | Approved, withdrawn |
| Boceprevir | Antiviral medication used as part of combination therapy in hepatitis C infection | Approved, withdrawn |
| Sodium aurothiomalate | Used for immunosuppressive anti-rheumatic effects | Approved, investigational |
| Felbinac | | Experimental |
| Fostamatinib | Indicated for treatment of rheumatoid arthritis and immune thrombocytopenic purpura | Approved, investigational |
| Cysteinesulfonic acid | | Experimental |

A search of https://clinicaltrials.gov and https://covid-trials.org found no trials related to COVID-19 for the drugs listed in this table.

Previous studies have suggested that an already approved drug, camostat mesylate, could be used as a TMPRSS2 inhibitor [74]. In this study, through rational querying of the knowledge graph, drug candidates for putative cathepsin L inhibitors have been identified (Table 2). These candidates are all approved drugs. Two of these drugs, telaprevir and boceprevir, have been used previously to help treat chronic hepatitis C infections; however, both have since been discontinued due to common side effects such as severe rashes, anemia, decreased neutrophils, and fatigue. Sodium aurothiomalate is a gold-containing compound approved for use treating rheumatoid arthritis. This drug is not currently commonly sold due to a difficulty in sourcing sodium aurothiomalate. The final potential cathepsin L inhibitor identified in this study is fostamatinib, which is approved for use against chronic immune thrombocytopenia. Other drug repurposing studies for COVID-19 have also identified this drug as a potential candidate, but as an ACE2 inhibitor [75]. These same studies, however, did not evaluate it as a cathepsin L inhibitor. Based on this and previous studies, it is therefore possible that a combined therapy of camostat mesylate and fostamatinib may aid in treating COVID-19.

Interestingly, fostamatinib was separately identified as a drug candidate via the PFIN clustering approach (Fig 3). Fostamatinib has serine/threonine-protein kinase TBK1 (Q9UHD2) as a drug target, which was also identified by Gordon *et al.* [27] as being a protein that interacts with a SARS-CoV-2 protein. However, the relevance of this interaction, if any, is unclear to us.

**Fig 3. Network view of a PFIN cluster containing fostamatinib.** Fostamatinib (highlighted in yellow) is identified as targeting a *Homo sapiens* protein, Q9UHD2 (red). Q9UHD2 is a high-confidence interactor of the SARS-CoV2 nsp13 protein.

## Potential antiviral modulators of inflammation

Severe COVID-19 disease often includes an exaggerated inflammatory response. Stebbing *et al.* suggested that combining antiviral and anti-inflammatory treatments is a possible means of reducing disease severity [76]. In their study, they examined the affinity between antiviral drugs (in particular, through inhibition of numb-associated kinases (NAKs)) and drug targets that may be useful in attenuating inflammation (e.g. through inhibition of Janus Kinases (JAKs)). As a result of the study, Stebbing *et al.* identified and discussed a range of drugs, including baricitinib, tofacitinib, and ruxolitinib.

To determine whether there were any additional drug candidates not discussed by Stebbing *et al.* [76], we queried our dataset for drugs targeting the NAKs and JAKs identified in the paper. These proteins were as follows: AAK1 (Q2M2I8), BIKE (Q9NSY1), GAK (O14976), JAK1 (P23458), JAK2 (O60674), JAK3 (P52333), and TYK2 (P29597). In addition to baricitinib, tofacitinib, and ruxolitinib, we also identified fostamatinib as a potential drug candidate. In fact, in our knowledge graph, fostamatinib was the only drug recorded as having all the aforementioned proteins as drug targets.

## Epidermal growth factor receptor (EGFR) inhibitors

EGFR (P00533) is a protein implicated in tissue fibrosis, due to its role in TGF-β1 dependent fibroblast-myofibroblast differentiation. Venkataraman and Frieman, 2017 [77], suggested that inhibiting EGFR signalling prevents excessive fibrotic responses and, thus, lung damage, during SARS infections.

Running the query in Fig 4 identified eight approved drugs that were possible EGFR inhibitors -- brigatinib, afatinib, osimertinib, fostamatinib, dacomitinib, neratinib, vandetanib, and panitumumab.

**Fig 4. Neo4j query used to identify EGFR inhibitors.** This query returns approved drugs that are annotated as targeting EGFR with an inhibitory or suppressive action.

## Lobeline, Nicotine and Galantamine

Two COVID proteins - ORF9c and NSP4 - have interactions with mitochondria-related human proteins ZNT6 (Q6NXT4, encoded by the SLC30A6 gene, transmembrane zinc transporter located in the golgi apparatus) and TIM29 (Q9BSF4, encoded by the TIMM29 gene, inner mitochondrial membrane translocase), respectively [27]. The genes encoding these proteins were both identified as having a high-confidence functional interaction with the neuronal acetylcholine receptor subunit alpha-9 (NACHR9) protein (Q9UGM1, gene name CHRNA9) [78,79]. These relationships are shown in Fig 5. NACHR9 is a is part of the ligand-gated ionic channel and nicotinic acetylcholine receptor superfamilies, and forms homo- or hetero-oligomeric divalent cation channels in the plasma membrane [80].

**Fig 5. Portion of one of the PFIN clusters, showing the local neighbourhood around the human protein NACHR9.** Lobeline can be seen to target the protein encoded by CHRNA9. The two human proteins (Q6NXT4 and Q9BSF4) shown to interact with SARS-CoV-2 proteins are highlighted in red. Q6NXT4 is a high-confidence interactor of the SARS-CoV-2 orf9c and Q9BSF4 is a high-confidence interactor of the SARS-COV-2 nsp4.

According to DrugBank, NACHR9 is targeted by the following small molecules: lobeline; galantamine; nicotine; tetraethylammonium; RPI-78M and, ATG003. Lobeline, in its natural form from plants in the *Lobelia* genus, has been proposed and applied for therapeutic uses including respiratory disorders (such as asthma) as a stimulant to treat wheezing, uncontrollable coughing and chest tightness [81], and has been reported to improve acute lung injury in cell lines [82]. There have been promising studies carried out on mice where lobeline helped to treat acute lung injury [83], but human studies are still required. Nicotine induces ACE2 overexpression in Human Bronchial Epithelial Cells (HBEpC) via alpha7-nicotinic receptor (α7-nAChR)[84], a paralog of NACHR9. Galantamine is a reversible, competitive inhibitor of acetylcholinesterase (AChE), which catalyses the breakdown of ACE, and is an allosteric modulator of nAChRs. It is currently approved for mild to moderate dementia and Alzheimer's.

## Quercetin

Quercetin is a polyphenol flavonoid found in many plants [85] and has been reported as having anti-inflammatory and anti-viral properties [86,87]. In one of the PFIN clusters, quercetin was identified as targeting the human protein casein kinase II subunit beta (CK2-beta) protein (P67870, localised throughout the cell), which Gordon *et al.* reported as being a high-confidence interactor of the SARS-CoV-2 N protein [27]. Two CK2-beta molecules, together with a single alpha and alpha' subunit together form the tetramer casein kinases II (CK2). CK2 acts as a serine/threonine-selective protein kinase, involved in cell cycle regulation, and is implicated in disorders of cell proliferation (e.g. tumours) [88]. Interestingly, CK2 has been found to be stimulated in other viral infections [89], and has various roles in the infection cycles of different viruses [90,91]. Further, Gordon *et al.* suggested CK2 inhibition as a possible therapeutic approach for COVID-19 [60], as CK2 downregulates stress granule formation [92] which is associated with enhanced viral replication in other coronavirus infections [93]. Thus, depending on the nature of the interaction between quercetin and CK2, it may be of benefit.

Quercetin has already been suggested as a possible therapeutic to mitigate the severity of SARS-CoV-2 infection, as quercetin alters the expression of 30% of the human proteins that are targets of SARS-CoV-2 proteins [94].

# Discussion

In summary, this study aimed to build on existing studies that have used knowledge frameworks to generate hypotheses about drug repurposing opportunities for COVID-19. As in other studies, an integrated dataset was developed, extending a drug repurposing framework developed by Mullen *et al.* [47]. We aimed to enhance the knowledge network approach through the use of PFINs to act as a framework to guide the hypothesis generation process. The use of PFINs allowed us to scale the data curation process in a systematic fashion, exploiting the combined effort of multiple curators. Our approach could act as a template for future studies, with richer knowledge graphs, the incorporation of algorithmic predictions for drug repurposing and a much larger set of curators. The drugs we identified for repurposing are illustrative of the value of our approach. Due to the rapidly developing nature of studies on SARS-CoV-2 and COVID-19, it is difficult to identify which candidates are truly novel, however, even in instances where candidates have already been suggested by other studies, our approach can still provide additional supporting (or refuting) evidence.

The results presented in this study are grouped by drug class. Where possible, we have tried to avoid suggesting specific instances of drugs (and focused on classes instead), in part because we recognise that our database is not complete. Further, ethically, there may be socioeconomic reasons for selecting one member of a drug class over another, such as supply and demand or the dependence of a subpopulation on a given drug. However, there are some noteworthy cases where focusing on specific drugs within a class is necessary. For example, fostamatinib has emerged as a drug candidate in a number of different contexts as an anti-inflammatory agent, as an EGFR inhibitor, and as a cathepsin L inhibitor, establishing it as a promising candidate.

One limitation of *in silico* drug repurposing studies is that, depending on the algorithms employed, many drug repurposing candidates can be suggested. Refinement of a large set of drug candidates often requires many person-hours of time to curate, particularly when including time taken to understand the mechanisms / pathways involved. By (1) leveraging a PFIN-based clustering approach and (2) focusing on clusters containing high-confidence interactors with SARS-CoV-2 proteins, we were able to reduce the drug candidate search space significantly. Using a crowd-sourcing approach to analyse the clusters allowed individuals to curate sets of drug candidates that were likely to be mechanistically related, providing a logical partitioning of work. Curation allowed us to focus on drugs with additional context (e.g., pharmacological, biological) evidence to support or refute the drugs discussed, thus providing a balanced perspective on the drugs discussed. In future studies, it might be interesting to research systematic and computational approaches to integrating and ranking the findings of curators, perhaps building on larger systems for data sharing and integration such as FAIRDOMHub [95].

There are a number of caveats to this research which need to be addressed to ensure results are taken in an appropriate context. Firstly, the data sources contributing to our integrated semantic network are limited in scope and could be expanded. For example, the graph includes connections between drugs and their protein targets, but the nature (e.g. "inhibitory", "stimulatory", "suppressor") is not always recorded. Secondly, computational prediction can only ever generate suggestions or hypotheses based on the analysis of the limited data available. Thus, for any drug repurposing candidates suggested, it is critical that these are manually curated by experts in medicine, pharmacology, and human biology. Due to recent concerns over the potential for harm in COVID-19 preprints, we want to stress that this research should not be used to inform clinical practice, nor should people change their behaviour on the basis of it [96].

# Acknowledgements

The authors gratefully acknowledge contributions from the Interdisciplinary Computing and Complex Biosystems research group at the University of Newcastle upon Tyne.

# Supporting information

**S1 Table. Existing approved drugs that inhibit CALM1.** These drugs have a range of indications, from psychiatric disorders to cardiovascular disorders.

| Drug | Indication (if recorded in database) |
| --- | --- |
| Phenoxybenzamine | |
| Chlorpromazine | Psychotic disorder, acute intermittent porphyria, schizophrenia, manic bipolar affective disorder |
| Nifedipine | Hypertensive disorder, prinzmetal angina |
| Pimozide | Tourette syndrome |
| Promethazine | Atopic conjunctivitis, urticaria, vasomotor rhinitis, seasonal allergic rhinitis |
| Perphenazine | Mixed anxiety and depressive disorder, schizophrenia |
| Loperamide | |
| Trifluoperazine | Schizophrenia |
| Fluphenazine | Schizophrenia, psychotic disorder |
| Cinchocaine | |

# Figures

Figure 1

```
UNWIND {gene_identifiers} as i

MATCH (gene:Gene {primaryDomainId:i})

OPTIONAL MATCH (gene)<-[peg:ProteinEncodedBy]-(pro:Protein)

OPTIONAL MATCH (pro)<-[dht:DrugHasTarget]-(drug)

OPTIONAL MATCH (drug)-[dsim:MoleculeSimilarityMolecule]-(drug1)

WHERE dsim.morganR2 > 0.5

OPTIONAL MATCH (gene)-[gawd:GeneAssociatedWithDisorder]-(disorder)

RETURN gene, peg, pro, drug, disorder, dht, gawd, dsim, drug1
```

Figure 2

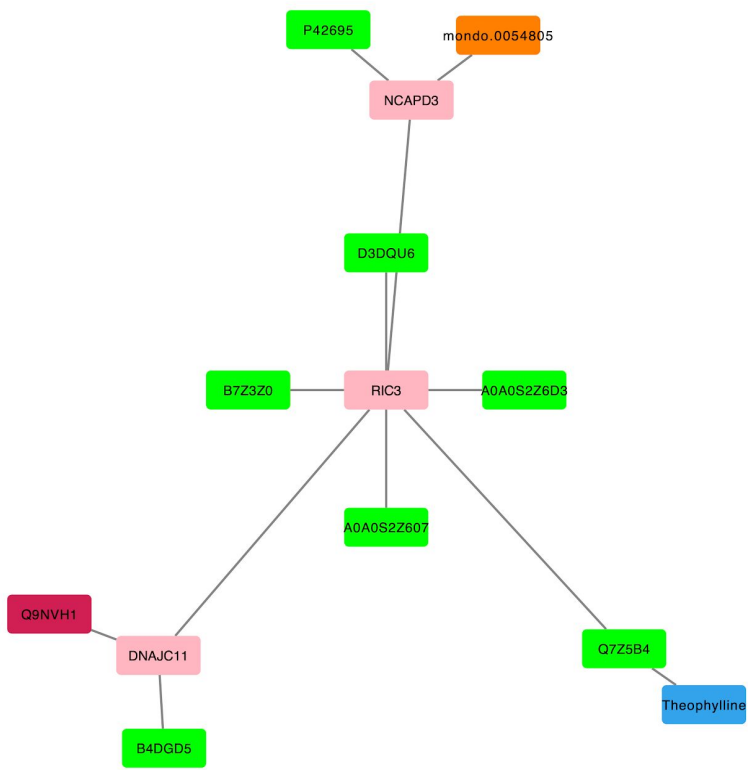

Figure 3

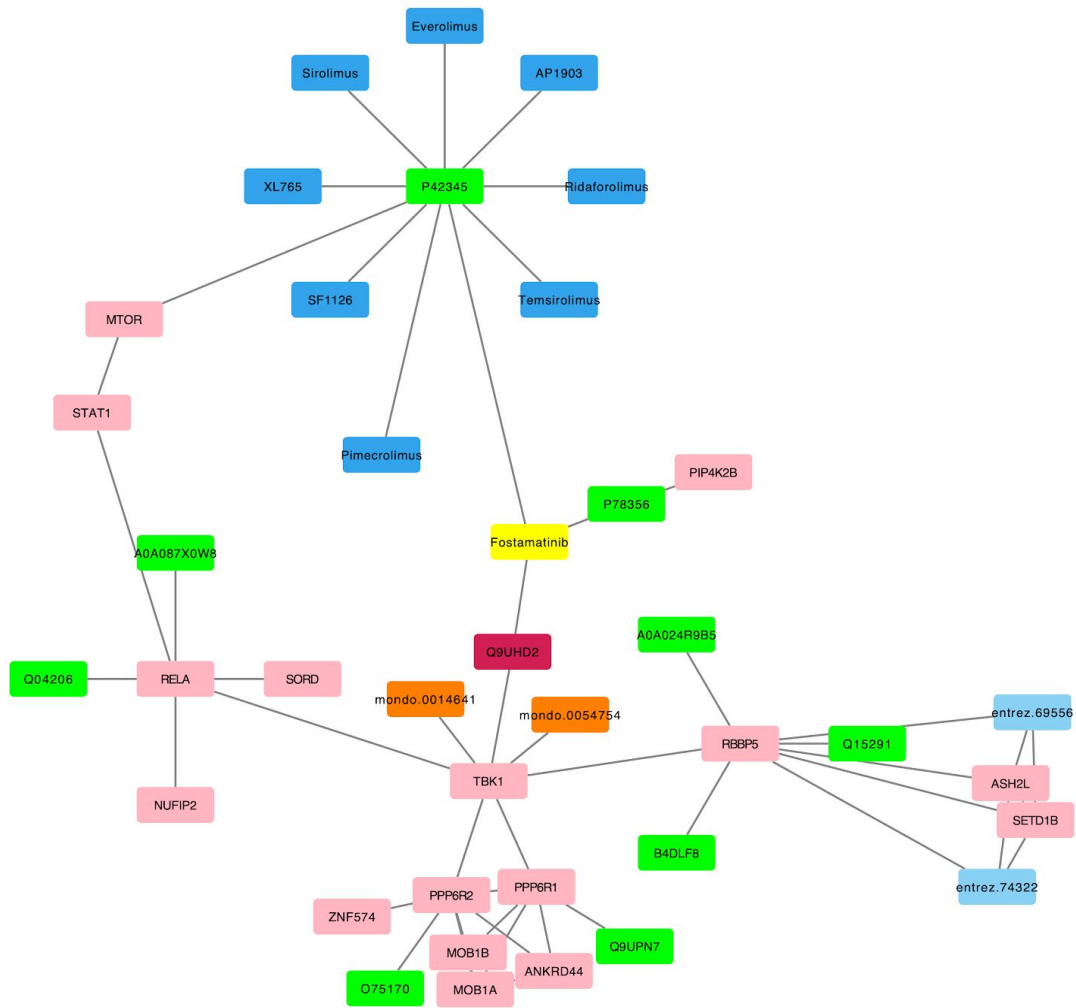

Figure 4

```
MATCH (n:Protein {primaryDomainId:"uniprot.P00533"})<-
[dt:DrugHasTarget]-(drug)
WHERE ("approved" in drug.drugGroups) AND ("inhibitor" in
dt.actions OR "suppressor" in dt.actions)
RETURN drug, n
```

Figure 5

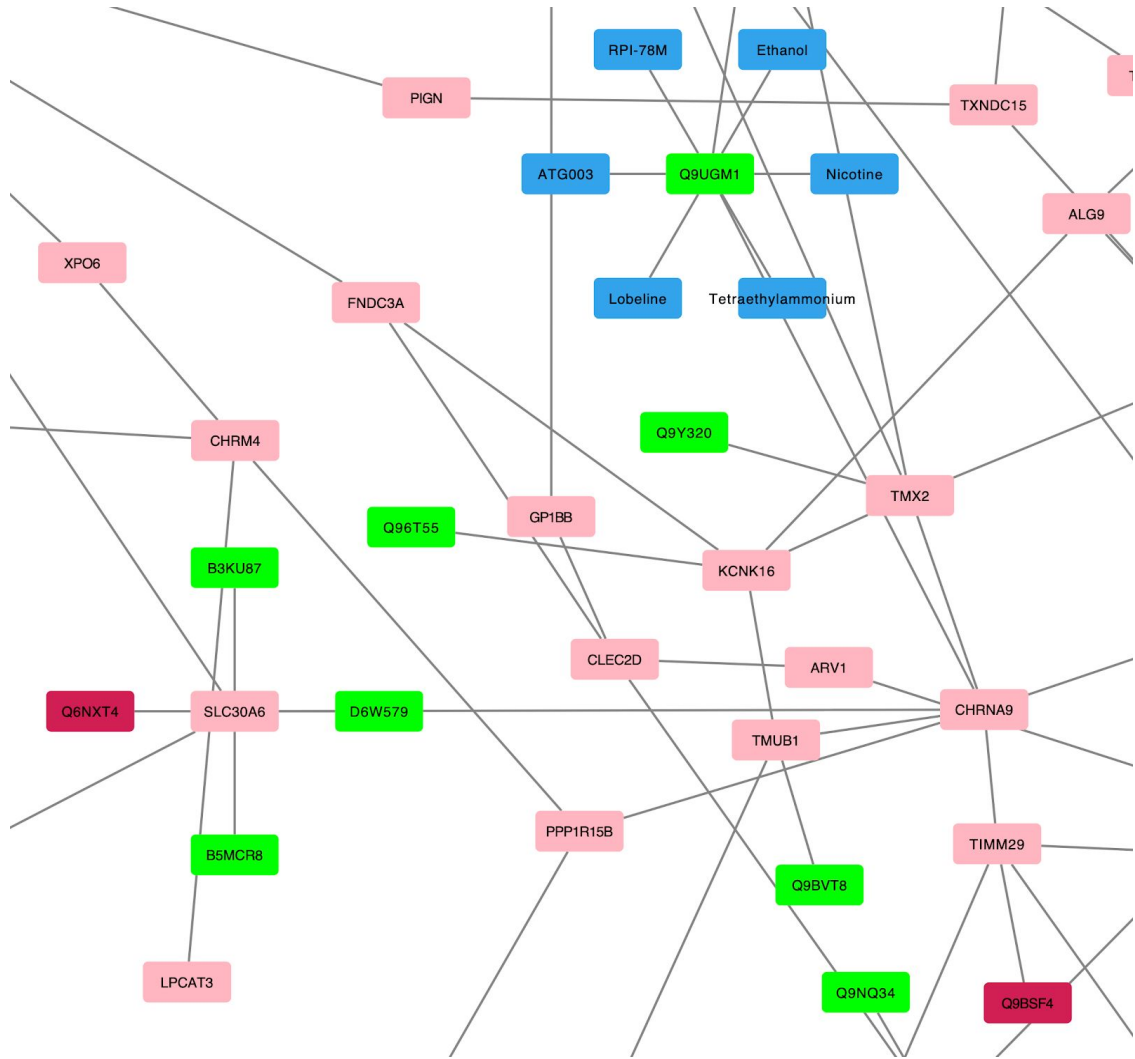